\documentclass{article}

\usepackage{arxiv}

\usepackage{amssymb}
\usepackage{amsmath,amsfonts}
\usepackage{algorithmic}
\usepackage{algorithm}
\usepackage{array}
\usepackage[caption=false,font=normalsize,labelfont=sf,textfont=sf]{subfig}
\usepackage{textcomp}
\usepackage{stfloats}
\usepackage{url}
\usepackage{verbatim}
\usepackage{graphicx}
\usepackage{cite}
\usepackage[utf8]{inputenc} 
\usepackage[T1]{fontenc}    
\usepackage{hyperref}       
\usepackage{url}            
\usepackage{booktabs}       
\usepackage{amsfonts}       
\usepackage{nicefrac}       
\usepackage{microtype}      
\usepackage{lipsum}
\usepackage{changepage}
\usepackage{graphicx}
\usepackage{amsmath}
\usepackage{amsfonts}
\usepackage{amssymb}
\usepackage{gensymb}
\usepackage{makeidx}
\usepackage{algorithm,algorithmic}
\usepackage[switch]{lineno}
\usepackage{xcolor}
\usepackage{bm}
\usepackage[justification=centering]{caption}
\usepackage{amsfonts}
\usepackage[bbgreekl]{mathbbol}
\DeclareSymbolFontAlphabet{\mathbbm}{bbold}
\DeclareSymbolFontAlphabet{\mathbb}{AMSb}%
\usepackage{amsthm}
\newtheorem{theorem}{Theorem}
\newtheorem{problem}{Problem}

\newtheorem{lemma}{Lemma}
\usepackage{natbib}
\usepackage[normalem]{ulem}

\title{Discovering Nonlinear Static Relationships in\\  Unlabeled Dataset using\\ Autoencoder with Ordered Variance}

\author{
  Midhun T. Augustine, Parag Patil, Mani Bhushan, and Sharad Bhartiya\\Automation Lab, 
  Department of Chemical Engineering\\
  Indian Institute of Technology Bombay, India\\
  \texttt{\{30004946,30004930,mbhushan,sharad\_bhartiya\}@iitb.ac.in} \\
  \vspace{.01cm}\\
  Date of initial version: 20 - 02 - 2024\\ 
   Date of current version: 30 - 05 - 2026\\\vspace{.01cm}
  }

\begin{document}
\maketitle

\begin{abstract}
This paper presents an autoencoder with ordered variance (AEO), in which the conventional reconstruction loss is augmented by a variance-based regularization term that promotes an ordered structure within the latent space. In this structure, the latent variables are ordered by their variance computed over the training data, facilitating systematic determination of the latent space dimensionality. The AEO is further extended using residual networks, resulting in a ResNet-based AEO (RAEO). Both AEO and RAEO green lead to discovery of nonlinear relationships among variables in unlabeled datasets, thereby enabling unsupervised static model extraction. Theoretical contributions include formal guarantees on the ordering of latent variances. The practical utility of the framework is demonstrated through its application to the identification of nonlinear steady-state models and their use in real-time optimization, with a continuous stirred tank reactor process serving as a representative case study. 

\end{abstract}


\section{Introduction}
\label{sec:sample1}

The main focus of this paper is on unsupervised  nonlinear static model extraction, which attempts to solve the following problem:
\begin{problem}[\textbf{Model extraction}]
 Given an unlabeled dataset containing $n$ variables $x_{1},x_{2},...,x_{n}$ with $p$ independent variables, and $n-p$ nonlinear relationships amongst the variables, find a model:
    \begin{equation}
        \label{eqmodel}
        \mathbf{f}(\mathbf{x})=\mathbf{0}
    \end{equation}
where $\mathbf{x}=\left[\begin{matrix} x_{1}\\  \vdots \\ x_{n}\end{matrix}\right] \in \mathbb{R}^{n},$ $\mathbf{f}:\mathbb{R}^{n} \rightarrow \mathbb{R}^{n-p},$ $p$ and $\mathbf{f}$ are unknown.
\end{problem}

The motivation for Problem 1 comes from the unsupervised learning hypothesis, which states that
real-world high-dimensional data are likely to concentrate in the vicinity of lower-dimensional manifolds due to system constraints \cite{Cayton2005}. This has led to the development of a subfield within unsupervised learning known as dimensionality reduction. 
Two central problems in this field are: (i) \emph{model extraction}, which seeks to uncover dependencies or relationships among the variables within a dataset, (ii) \emph{feature extraction}, which focuses on deriving lower-dimensional features from high-dimensional data. In cases where the relationships among variables in Eq. (\ref{eqmodel}) are linear, both model and feature extraction problems can be simultaneously solved using principal component analysis (PCA) \cite{Pearson1901,Jackson1991,Narasimhan2008}.  
In PCA, the objective is to identify an optimal linear transformation that projects high-dimensional, unlabeled data vectors onto a lower-dimensional latent feature space.
Even though PCA identifies and eliminates collinearity in the latent space, this ability degrades when the functional dependence in the variables becomes nonlinear.
  This has led to the introduction of nonlinear dimensionality reduction methods such as autoencoder (AE) \cite{Kramer1989}, principal curve \cite{Hastie1989}, kernel PCA \cite{Scholkopf1998}, isomap \cite{Tenenbaum2000}, diffusion map \cite{Nadler2005}, etc. Amongst these, AEs are popular because of their connection to neural networks and deep learning \cite{Li2022}. AEs can be considered as the nonlinear counterpart of PCA. However, in contrast to PCA, the latent variables in AEs are neither ordered nor guaranteed to be uncorrelated. Consequently, although AEs offer a viable approach to feature extraction, their inability to determine an optimal size or dimensionality for the latent space implies that they do not effectively address the model extraction problem.
  To address these challenges, various modifications of AEs are presented in the literature, which are mostly based on modifying the loss function with different regularization terms \cite{Kingma2014}-\cite{Wickramasinghe2021}.  
 However, these methods focus on feature extraction, and Problem 1 pertaining to nonlinear model extraction is not explored in any of these works. 
 
\par To summarize, despite its importance, Problem 1 on nonlinear model extraction from unlabeled dataset has not been explored in the existing literature to the best of the authors’ knowledge. This serves as the primary motivation for the present work, in which we propose a systematic approach for unsupervised nonlinear model extraction. To this end, we introduce an autoencoder with ordered variance (AEO) and its variant, ResNet AEO (RAEO).  The proposed AEO and RAEO with a regularized loss function result in the ordering of latent variables in terms of decreasing variance. These methods facilitate the discovery of nonlinear relationships among the variables in an implicit form as in Eq. (\ref{eqmodel}). Moreover, RAEO enables the extraction of explicit nonlinear relationships when dependent variables are specified by the user. Theoretical guarantees are provided for variance ordering in the AEO and RAEO framework. The codes for the numerical simulations presented in this paper are available on GitHub\footnote{\url{https://github.com/MIDHUNTA30/AEO}}.

\par \textit{Notations:} The $n$-dimensional Euclidean space is denoted by 
$\mathbb{R}^{n}$ and the space of $m \times n$ real matrices  by $\mathbb{R}^{m \times n}$.  
For a random vector   $\mathbf{x}\in \mathbb{R}^{n}$, the sample mean vector and sample covariance matrix are denoted by $\overline{\mathbf{x}}, \mathbf{V}_{\mathbf{x}}$, respectively. 
For a matrix $\mathbf{A}\in \mathbb{R}^{m \times n},$ the Frobenious norm is denoted as ${\parallel \mathbf{A} \parallel}_{F},$ where ${\parallel \mathbf{A} \parallel}_{F}=\sqrt{\sum_{i=1}^{m}\sum_{j=1}^{n}a_{ij}^{2}} = \sqrt{Trace(\mathbf{A}^{\top}\mathbf{A})}.$
For $\mathbf{A}\in \mathbb{R}^{m \times n},$ the $(i,j)$-th element is denoted by $\mathbf{A}_{(i,j)}.$ 
 The identity matrix of order $n\times n$ is represented by  $\mathbf{I}_{n}.$ An explicit model  is represented as $\mathbf{x}_{d}=\mathbf{f}_{ex}(\mathbf{x}_{i})$ and an implicit model as $\mathbf{f}(\mathbf{x}_{d},\mathbf{x}_{i})=\mathbf{0},$ where $\mathbf{x}_{d},\mathbf{x}_{i}$ denote the dependent and independent variables, respectively.

\section{Model extraction: Autoencoder with ordered variance (AEO)}
In this section, we introduce the AEO, which integrates a variance-based regularization term into the loss function to impose an ordered structure within the latent space. Given the unlabeled data vector $\mathbf{x}\in \mathbb{R}^{n}$ as the AE input, the encoder and decoder functions for the AEO are defined as follows:
\begin{equation}
  \label{eq4c1}   
 \begin{aligned} &\mathbf{y} =\mathbf{e} (\mathbf{x})=\mathbf{f}_{E}(\mathbf{f}_{{E-1}}(...(\mathbf{f}_{1}(\mathbf{x})))) \\
&\widehat{\mathbf{x}} =\mathbf{d} (\mathbf{y} )=\mathbf{f}_{{E+D}}(\mathbf{f}_{{E+D-1}}(...(\mathbf{f}_{{E+1}}(\mathbf{y} ))))
 \end{aligned}
\end{equation}
where  $\mathbf{y} \in \mathbb{R}^{m}$   is the latent vector, $\widehat{\mathbf{x}}\in \mathbb{R}^{n}$ is the reconstructed data vector, $\mathbf{e}:\mathbb{R}^{n} \rightarrow \mathbb{R}^{m}$ is the encoder subnetwork, $\mathbf{d}:\mathbb{R}^{m} \rightarrow \mathbb{R}^{n}$ is the decoder subnetwork,
$\mathbf{f}_{1},...,\mathbf{f}_{E}$ are layers of the encoder,  $\mathbf{f}_{{E+1}},...,\mathbf{f}_{{E+D}}$ are layers of the decoder, and
 $E$ and $D$ are the number of layers in the encoder and decoder, respectively. The layers $\mathbf{f}_{i}$  can be represented as:
\begin{equation}
   \mathbf{f}_{i}(\cdot)=\bm{\sigma}_{i}(\mathbf{A}_{i}(\cdot)+\mathbf{b}_{i})
\end{equation}
where  $\bm{\sigma}_{i}: \mathbb{R}^{m_i} \rightarrow \mathbb{R}^{m_i}$ contains the element-wise activation functions for the $i^{th}$ layer consisting of $m_{i}$ neurons, $\mathbf{A}_{i} \in \mathbb{R}^{m_{i} \times m_{{i-1}} }$ is the weight matrix, $\mathbf{b}_{i} \in \mathbb{R}^{m_{i}}$ is the bias vector,  
 $i=1,2,...,E+D$,  $m_{0}=n,m_{E}=m,$ and $ m_{{E+D}}=n$.

Let $\bm{\theta}$ denote the set of all trainable parameters comprising weights and biases across all layers of the encoder and decoder subnetworks. Define the AE training, latent, and reconstructed data matrices, respectively, as:
  \begin{equation}
      \label{eq1z}  
      \begin{aligned}
      &\mathbf{X}=\left[\begin{matrix} \mathbf{x}_{1} & \mathbf{x}_{2} & \dots & \mathbf{x}_{N}\end{matrix}\right] \in \mathbb{R}^{n \times N}\\
      &\mathbf{Y}=\left[\begin{matrix} \mathbf{y}_{1} & \mathbf{y}_{2} & \dots & \mathbf{y}_{N}\end{matrix}\right]\in \mathbb{R}^{m \times N}\\ &\widehat{\mathbf{X}}=\left[\begin{matrix} \widehat{\mathbf{x}}_{1} & \widehat{\mathbf{x}}_{2} & \dots & \widehat{\mathbf{x}}_{N}\end{matrix}\right]\in \mathbb{R}^{n \times N}
      \end{aligned}
    \end{equation}
where $N$ is the number of samples, and the training data is assumed to be mean-centered.
The loss function for the proposed AEO is defined as:
\begin{equation}
 \label{eq3a} 
J = { \underbrace{{\parallel \mathbf{X}-\widehat{\mathbf{X}}  \parallel }_{F}^{2}}_{J_{x}}+ \alpha\underbrace{  {\parallel \mathbf{Q}^{\frac{1}{2}}[\mathbf{Y} -\overline{\mathbf{Y}} ] \parallel }_{F}^{2}}_{J_{y}} + \beta \underbrace{ {\parallel \bm{\theta}   \parallel }_{2}^{2}}_{J_{\theta}}}
\end{equation}
where $\overline{\mathbf{Y}} =\left[\begin{matrix} \overline{\mathbf{y}}  & \overline{\mathbf{y}}  & \dots & \overline{\mathbf{y}}   \end{matrix}\right]\in \mathbb{R}^{m  \times N}$ contains the mean latent vector $\overline{\mathbf{y}} $ as its columns, $\mathbf{Q}=diag(q_{1},q_{2},...,q_{m})$  is the weighting matrix, $J_{x}$ is the reconstruction error term,  $J_{y}$ is the proposed variance regularization term,  $J_{\theta}$ is the weight regularization term, and $\alpha,\beta>0$ are hyperparameters. The elements of the weighting matrix are selected in strictly increasing order, satisfying $0\leq q_{1} <q_{2}<\dots<q_{m}< \infty$.   In this paper, an exponential weighting scheme is adopted, where each element is defined as: $q_{j}=c^{j},$ $j=1,2,\dots,m$  with $c>1$ being a hyperparameter. Other choices, such as linear, polynomial, and logarithmic growth of the $q_j$ may also be employed. 
The optimization problem for AEO becomes:
\begin{equation}
\label{eq3y}
    \underset{\bm{\theta} }{\min} \hspace{.2cm} J
\end{equation}
where $J$ is the loss function in Eq. (\ref{eq3a}).
 An approach to extract a nonlinear model using AEO is presented next.

\subsection{Extracting nonlinear relationships using AEO}
\begin{figure*}[h]
 		\begin{center}
 		\includegraphics [scale=.825] {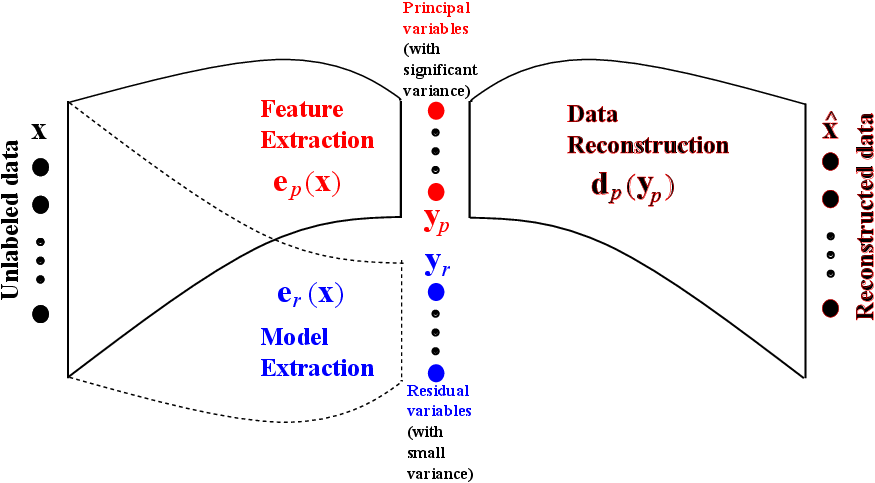}
 		\caption{{Unsupervised model extraction using AEO schematic representation.}}
 		\label{fig4umAEO}		
 	\end{center}
 \end{figure*}

This section discusses the application of the AEO framework to nonlinear model extraction as stated in Problem 1.  
Without loss of generality, the training data is 
assumed to be mean-centered.  Additionally, the activation functions for the output layers of the encoder and decoder, denoted by $\bm{\sigma}_{E}$ and $\bm{\sigma}_{E+D}$ respectively, are chosen to be linear, for which the bias terms are chosen as zero, consistent with the mean-centered nature of the data. This simplifies $\mathbf{y} $ and $\widehat{\mathbf{x}} $ in Eq. (\ref{eq4c1}) as
\begin{equation}   
\label{eq3p1} 
\begin{aligned}
&\mathbf{y} =\mathbf{A}_{E}\mathbf{f}_{{E-1}}(...(\mathbf{f}_{{1}}(\mathbf{x})))\\
&\widehat{\mathbf{x}} =\mathbf{A}_{{E+D}}\mathbf{f}_{{E+D-1}}(...(\mathbf{f}_{{E+1}}(\mathbf{y} ))).
\end{aligned}
\end{equation}
Analogous to PCA (see   \ref{secA1}), the proposed AEO facilitates the identification of latent variables with near-zero variance, which can be exploited to reveal underlying nonlinear relationships as in Eq. (\ref{eqmodel}). For this purpose, an AEO is trained with an initial latent dimensionality of $m = n,$ using the available training data. The goal is to identify latent variables $y_j$  with nearly zero variances, which is achieved by computing the reduced latent dimensionality $p$ as: 
 \begin{equation}
\label{equation_latent_dimensionality}
\begin{aligned}
p=\underset{j\in\{1,2,\dots,m\}}{\mathrm{argmax}} \hspace{.2cm}  &
j \\
\mathrm{subject \hspace{0.1cm} to}~&~ V_{y_j} \geq \delta.
\end{aligned}
 \end{equation}
 where $\delta > 0$ is a user-defined threshold and $V_{y_j}$ is the variance of the $j^{th}$ latent variable: $V_{y_j}=\mathbf{V}_{\textbf{y}}(j,j)$. 
Since the variance of latent variables in AEO is ordered, the first $p$ latent variables which exhibit significant variance ($V_{y_j} \geq \delta$) are treated as principal latent variables, and
 the latent variables from $p+1$ to $n$ with nearly zero variance are treated as residual latent variables. Thus, the elements of $\mathbf{y} ,$ 
 $\mathbf{A}_{E}$ and $\mathbf{A}_{{E+1}}$ in Eq. (\ref{eq3p1}) can be grouped as follows:
\begin{equation}
\label{eq3c} 
 {\mathbf{y}  =\left[\begin{matrix} \mathbf{y}_{p} \\ \mathbf{y}_{r} \end{matrix}\right]}, \hspace{.2cm} 
   {\mathbf{A}_{E} =\left[\begin{matrix} \mathbf{A}_{{E_{p}}} \\   \mathbf{A}_{{E_{r}}} \end{matrix}\right]}, \hspace{.2cm} 
   {\mathbf{A}_{{E+1}} =\left[\begin{matrix} \mathbf{A}_{{{E+1}_{p}}} &   \mathbf{A}_{{{E+1}_{r}}} \end{matrix}\right]}
\end{equation}
where $\mathbf{y}_{p} \in \mathbb{R}^{p}, \mathbf{y}_{r} \in \mathbb{R}^{n-p}, \mathbf{A}_{{E_p}} \in \mathbb{R}^{p \times m_{{E-1}}},$  $\mathbf{A}_{{E_r}} \in \mathbb{R}^{(n-p) \times m_{{E-1}}},$ $\mathbf{A}_{{{E+1}_p}} \in \mathbb{R}^{ m_{{E+1}} \times p}, $  $\mathbf{A}_{{{E+1}_r}} \in \mathbb{R}^{ m_{{E+1}} \times (n-p)}.$
Using these, the latent vector in Eq. (\ref{eq3p1}) can be rewritten as:
\begin{equation}
\label{eq3d} 
\mathbf{y} =\left[\begin{matrix} \mathbf{y}_{p} \\ \mathbf{y}_{r} \end{matrix}\right]=
\left[\begin{matrix} \mathbf{A}_{{E_{p}}}  \\  \mathbf{A}_{{E_{r}}} \end{matrix}\right]\mathbf{f}_{{E-1}}(...(\mathbf{f}_{1}(\mathbf{x}))).
\end{equation}
The part of the encoder function corresponding to the principal and residual variables is denoted by $\mathbf{e}_{p},\mathbf{e}_{r}$, respectively, i.e., $\mathbf{y}_{p}= \mathbf{e}_{p}(\mathbf{x})$ and $\mathbf{y}_{r}= \mathbf{e}_{r}(\mathbf{x})$. Given that the residual latent variables have small variances, they can be approximated as $\mathbf{y}_{r}=\overline{\mathbf{y}}_{r}$. 
Substituting this in Eq. (\ref{eq3d}) results in the $n-p$ nonlinear implicit relationships among the $n$ variables in $\mathbf{x}$ as:
\begin{equation}
\label{eq3e} 
\underbrace{\mathbf{A}_{{E_{r}}} \mathbf{f}_{{E-1}}(...(\mathbf{f}_{1}(\mathbf{x})))}_{\mathbf{e}_{r}(\mathbf{x})}-\overline{\mathbf{y}}_{r}=\mathbf{0}.
\end{equation}
Given an appropriate set of $p$ variables in $\mathbf{x},$ one can solve Eq. (\ref{eq3e}) for the remaining $n-p$ variables using numerical solvers for nonlinear system of equations \cite{Benton2018}.   A schematic representation of unsupervised model extraction using AEO is given in Fig. \ref{fig4umAEO}. The computation of the reconstructed vector can be simplified by replacing  $\mathbf{y}_{r}$ with $\overline{\mathbf{y}}_{r}$, which modifies the reconstructed vector $\widehat{\mathbf{x}} $ in Eq. (\ref{eq3p1}) as
\begin{equation}
\label{eqrec}
   \begin{aligned}
\widehat{\mathbf{x}} &=\mathbf{A}_{{E+D}}\mathbf{f}_{{E+D-1}}(...(\sigma_{{E+1}}(\mathbf{A}_{{{E+1}_{p}}} \mathbf{y}_{p} + \mathbf{A}_{{{E+1}_{r}}} \mathbf{y}_{r}+\mathbf{b}_{{E+1}})))\\
& \approx \underbrace{\mathbf{A}_{{E+D}}\mathbf{f}_{{E+D-1}}(...(\sigma_{{E+1}}(\mathbf{A}_{{{E+1}_{p}}} \mathbf{y}_{p} +\mathbf{b}_{\text{r}_{E+1}})))}_{\mathbf{d}_{p}(\mathbf{y}_{p})}
\end{aligned} 
\end{equation}
where $\mathbf{b}_{\text{r}_{E+1}}=\mathbf{A}_{{{E+1}_{r}}} \overline{\mathbf{y}}_{r}+\mathbf{b}_{{E+1}}$ and $\mathbf{d}_{p}:\mathbb{R}^{p} \rightarrow \mathbb{R}^{n}.$ Note that Eq. (\ref{eqrec}) gives the reconstructed vector in terms of the principal latent variables as illustrated in Fig. \ref{fig4umAEO}. The stepwise procedure for nonlinear model extraction using AEO is summarized in Algorithm 1.

\begin{algorithm}[H]
 
	\begin{algorithmic}[1] 	
	\STATE Given $\mathbf{X}$. 		 Select  $m=n, \delta,$ $\alpha,\beta,\mathbf{Q}$
	
		\STATE  Compute $\bm{\theta}^{*}$ by solving optimization problem in Eq. (\ref{eq3y})

        \STATE  Compute $\mathbf{Y}$ as in Eq. (\ref{eq1z}) for the AEO with the parameter $\bm{\theta}^{*}$
		\STATE Compute $\mathbf{V}_{\mathbf{y}}=\frac{1}{N-1}[\mathbf{Y}-\overline{\mathbf{Y}}][\mathbf{Y}-\overline{\mathbf{Y}}]^{\top}$
        \STATE   $i=m$
        \WHILE    {$V_{y_{i}} < \delta$}
        \STATE $ i-1 \leftarrow i$ 
        \ENDWHILE
         \STATE $p= i$
        \STATE Compute $\mathbf{A}_{E_p}$ and $\mathbf{A}_{E_r}$  using Eq. (\ref{eq3c})
        
        \STATE Obtain nonlinear model as in Eq. (\ref{eq3e}).
	\end{algorithmic}
	\caption{Model extraction using AEO}
\end{algorithm}

\subsection{Illustrating AEO}
\par This section illustrates the proposed AEO on a five-variable dataset in which the  variables are related by the following nonlinear equations:
\begin{equation}
\label{eq5a} 
   \begin{aligned}       
   & x_{4}-sin(3 x_{1}) =0\\
   &x_{5}-x_{2}+tan(0.5x_{3})=0.
       \end{aligned}
\end{equation}
To generate the data, $N=500$ samples of $x_{1},x_{2},x_{3}$ are selected, which are chosen as uniform random variables in the range of -1 to 1. From this, the samples for $x_{4}$ and $x_{5}$ are generated using Eq. (\ref{eq5a}) to which Gaussian noise with zero mean and standard deviation of $0.1$ is added. 
The first 300 samples of the dataset are used
for training AEO, while the remaining samples are used for 
 testing. The parameters for AEO are chosen as $E=2,D=2,$ $m_{1}=m_{3}=6,$ $m_{2}=m_{4}=n=5,$ $\alpha=0.001,$ $\beta=0.5$.
The weighting matrix elements for AEO are chosen as $q_{i}=c^i$ with $c=10$.   The variance of the latent variables computed over the training data are obtained as: $V_{y_1}=0.1626,$ $V_{y_2}=0.0648,$ $V_{y_3}=0.0143,$ $V_{y_4}=3.3 \times 10^{-7},$  $V_{y_5}=1.6\times 10^{-7}$  which gives the reduced latent dimensionality: $p=3$, for $\delta=0.001$. An implicit relationship between the variables $x_{1}$ to and $x_{5}$ can be obtained by substituting for $\mathbf{A}_{1},$ $\mathbf{A}_{{2_{r}}},$ and $\overline{\mathbf{y}}_{r}$ in Eq. (\ref{eq3e}) which gives:
\begin{equation}
\label{eqimpaeo}
\begin{aligned}
 {{ \left[\begin{matrix}0.0022  & 0.0970 \\ 0.0326 & 0.0565  \\  -0.0262 & -0.1186 \\  -0.0054 & 0.0085 \\ 0.0115 & -0.0172 \\ 0.0087 &  0.0216\end{matrix}\right]}}^{\top}\mathbf{tanh} \Bigg(   {{\left[\begin{matrix}  0.0386 &    0.0078 &    0.1223  & 0.0826 & 0.0625   \\   -0.0873 &  -0.0002 &   0.0893  & -0.0899 & 0.0819\\    -0.0436 &   0.0431 &    0.1101 & 0.0125 & 0.0897 \\
    -0.0045 & 0.1537 & -0.0822 & -0.1162 & 0.0761\\0.2250& 0.0121 & 0.0203 & 0.1002 & 0.0604 \\ -0.0085 & 0.1331 & -0.1419 & 0.0600 & 0.0361 \end{matrix}\right]}}{{\left[\begin{matrix} x_{1} \\ x_{2} \\ x_{3}\\x_{4} \\x_{5} \end{matrix}\right]}}
 \Bigg)\\ +   {{\left[\begin{matrix}   -0.0000006 \\ 0.000023 \end{matrix}\right]}} = {{\left[\begin{matrix}   0 \\ 0 \end{matrix}\right]}}
 \end{aligned}
\end{equation}
\begin{figure}[h]
 		\begin{center}
 		\includegraphics [scale=.475] {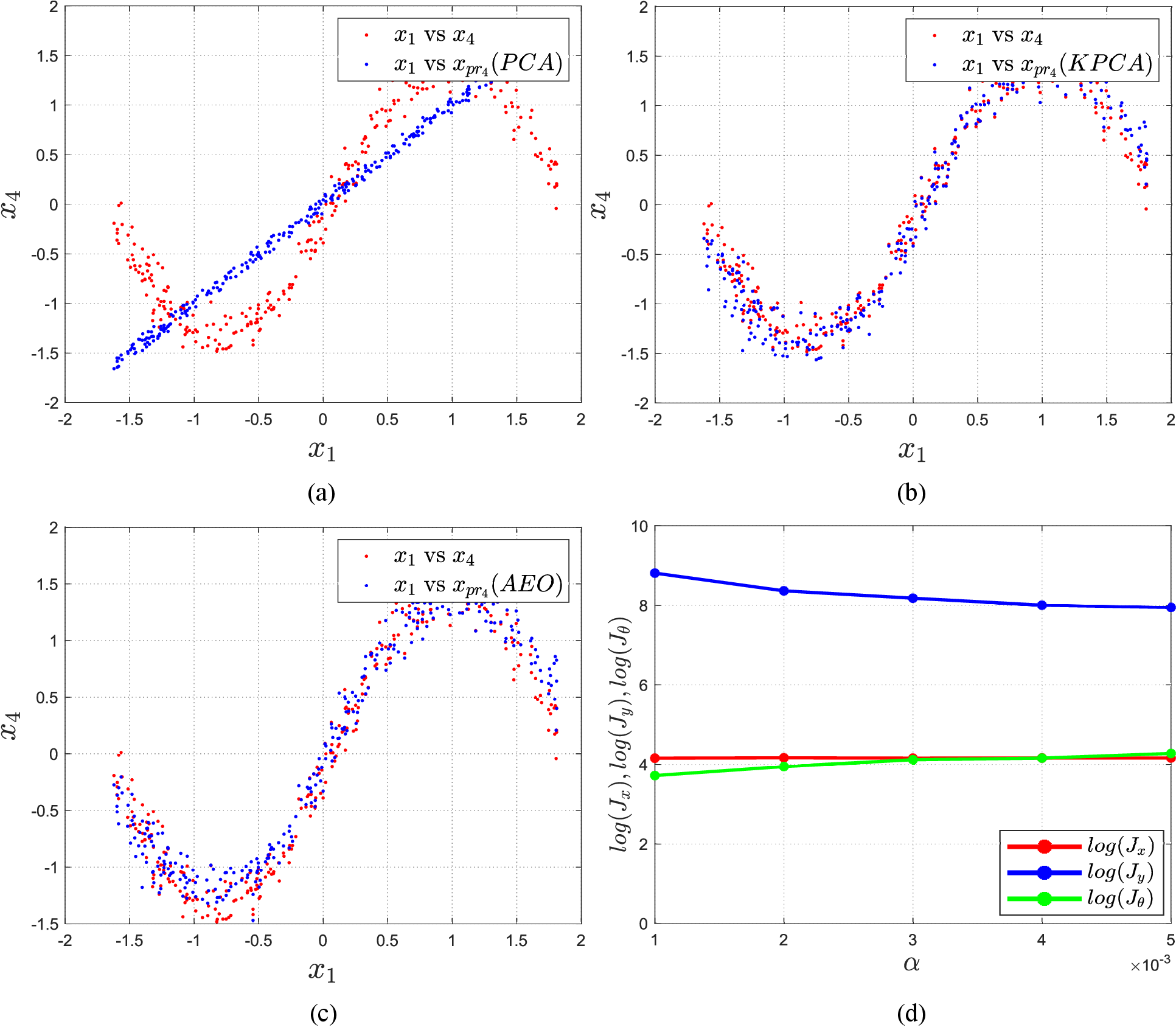}
 		\caption{{Model extraction with (a) PCA, (b) KPCA, (c) AEO, (d) AEO sensitivity analysis for  $\alpha$}}
 		\label{figcomp1}		
 	\end{center}
 \end{figure}
\par Fig. \ref{figcomp1}(a)-(c)  shows the plot of $x_{4}$ $vs$ $x_{1}$  for both the actual and predicted data (denoted by $x_{pr}$) with PCA, KPCA, and AEO, respectively. The PCA prediction is based on a linear model extracted as in  \ref{secA1}, the KPCA  uses nonlinear model extracted as in  \ref{secA2}, and the AEO uses nonlinear model in Eq. (\ref{eqimpaeo}). 
Comparing Fig. \ref{figcomp1}(a) with Figs. \ref{figcomp1}(b), (c) clearly indicates the superiority of AEO and KPCA over PCA. Here PCA, which gives a linear approximation,  fails to extract the sinusoidal relationship in the data. Whereas, both AEO and KPCA successfully capture the nonlinear relationships within the dataset. Note that the proposed AEO model is parametric in nature, whereas the KPCA model is nonparametric. Consequently, the complexity of the KPCA model depends on the training data itself rather than on a fixed set of parameters.
The major challenge associated with the AEO approach is the selection of hyperparameters, which must be chosen appropriately to ensure accurate model order extraction.  To simplify the process of hyperparameter selection, we can initially fix the values of $\beta$ and $c$, and then tune $\alpha$ to achieve the desired performance. 
Figure \ref{figcomp1}(d)  shows  the values of $J_{x},J_{y},J_{\theta}$ on a logarithmic scale for $\alpha$ ranging from 0.001 to 0.005 in steps of 0.001, with $\beta=0.5$ and $c=10$ held fixed. As $\alpha$ increases over this interval, $J_{y}$ decreases from $6692.8$ to $2823.8$, whereas $J_{x}$ increases slightly from $63.7036$ to $64.0143,$ and $J_{\theta}$ increases from $41.1311$ to $ 71.6708$. 
Generally, increasing the value of $\alpha$ enhances dimensionality reduction by driving more latent variables towards zero, which may, in turn, impact the reconstruction error. Consequently, $\alpha$ can be selected as a trade-off between reconstruction error and latent dimensionality.
 \par The nonlinear relationships extracted by AEO are in implicit form, as in Eq. (\ref{eqimpaeo}). Although these implicit relationships can be solved numerically using various solvers, the accuracy and reliability of the solutions depend on factors such as the selection of an initial guess, the existence and uniqueness of solutions, and the stability of the numerical method used. Further, obtaining an explicit relationship is beneficial in real-time applications such as data reconciliation \cite{Marimuthu2019}, soft sensing \cite{Shang2014}, process monitoring \cite{Lee2004,ZLi2022}, and real-time optimization \cite{Trierweiler2014}.  This motivates an extension of the AEO framework. In the following section, we introduce the RAEO, in which the encoder is implemented using a ResNet architecture to improve representational capacity. The section further includes a discussion on the extraction of explicit nonlinear models using the RAEO framework.

\section{Explicit model extraction: ResNet autoencoder with ordered variance (RAEO)}

\begin{figure*}[h]
 		\begin{center}
 		\includegraphics [scale=.625] {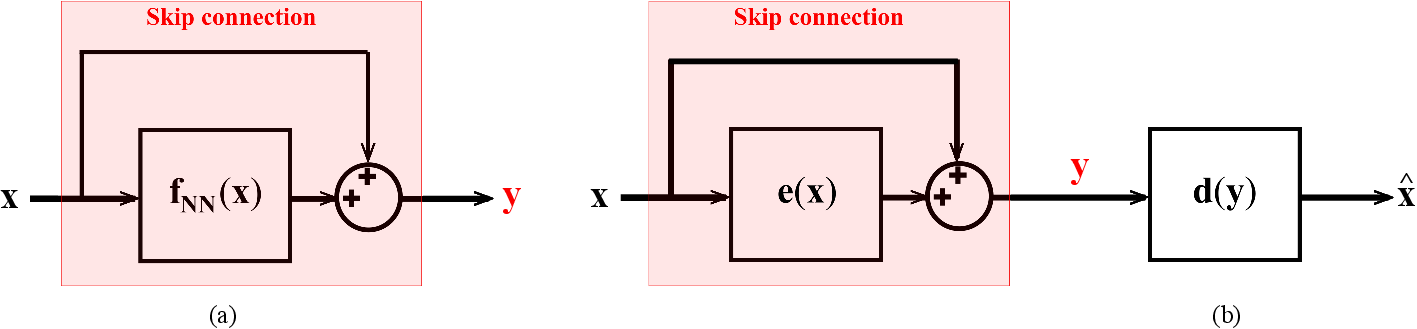}
 		\caption{{(a) ResNet (b) ResNet AEO.}}
 		\label{fig4a}		
 	\end{center}
 \end{figure*}

Residual neural networks or residual networks (ResNets) are neural networks that have a residual connection or skip connection, as shown in Fig. \ref{fig4a}. In general, a ResNet can be mathematically represented as \cite{He2016}:
\begin{equation}
\label{eq4res} 
  \mathbf{y} = \mathbf{x}+\mathbf{f}_{\text{NN}}(\mathbf{x})
\end{equation}
where $\mathbf{x}\in \mathbb{R}^{n}, \mathbf{y} \in \mathbb{R}^{n},$ and   $\mathbf{f}_{\mathbf{NN}}: \mathbb{R}^{ n} \rightarrow \mathbb{R}^{n}$ is a feedforward NN. In the proposed RAEO, the encoder function is chosen as a ResNet as in Eq. (\ref{eq4res}) and the decoder is chosen as an NN.
  This gives the latent and reconstructed vectors for RAEO as:
\begin{equation}   
\label{eq4c} 
\begin{aligned}
\mathbf{y}&= \mathbf{x}+\mathbf{e}(\mathbf{x})=\mathbf{x}+\mathbf{f}_{E}(\mathbf{f}_{{E-1}}(...(\mathbf{f}_{1}(\mathbf{x}))))\\
\widehat{\mathbf{x}}&=\mathbf{d}(\mathbf{y})= \mathbf{f}_{{E+D}}(\mathbf{f}_{{E+D-1}}(...(\mathbf{f}_{{E+1}}(\mathbf{y}))))
\end{aligned}
\end{equation} 
 The loss function for the RAEO 
 is chosen as in Eq. (\ref{eq3a}). 
Now, the optimization problem for RAEO can be defined as in Eq. (\ref{eq3y}).
 The application of RAEO in nonlinear
model extraction is discussed next.

\subsection{Extracting nonlinear relationships using RAEO}
  This section focuses on the extraction of explicit nonlinear models using the RAEO framework. To this end, RAEO is trained on the available data with the objective of identifying components of the latent vector $\mathbf{y}$ that exhibit very low  variance. Consistent with the assumptions made in the AEO framework, the training data is assumed to be mean-centered. Furthermore, the activation functions used in the output layers of both the encoder and decoder are linear, and the associated bias terms are set to zero. This simplifies $\mathbf{y}$ and $\widehat{\mathbf{x}}$ for RAEO as follows:
\begin{equation}   
\label{eq3p} 
\begin{aligned}
&\mathbf{y}=\mathbf{x}+\mathbf{A}_{E}\mathbf{f}_{{E-1}}(...(\mathbf{f}_{1}(\mathbf{x})))\\
&\widehat{\mathbf{x}}=\mathbf{A}_{{E+D}}\mathbf{f}_{{E+D-1}}(...(\mathbf{f}_{{E+1}}(\mathbf{y}))).
\end{aligned}
\end{equation}
  Let the 
  latent variables from $p+1$ to $n$ have small variance. Then elements in $\mathbf{y}$ and $\mathbf{A}_{E}$ can grouped as in Eq. (\ref{eq3c}). Further, the elements in $\mathbf{x}$ and $\mathbf{A}_{1}$ are grouped as:
  \begin{equation}
       \mathbf{x} =\left[\begin{matrix} \mathbf{x}_{p} \\ \mathbf{x}_{r} \end{matrix}\right], \hspace{.5cm} 
      \mathbf{A}_{1} =\left[\begin{matrix} \mathbf{A}_{1_{p}} &  \mathbf{A}_{1_{r}}  \end{matrix}\right]
  \end{equation}
  where $\mathbf{x}_{p} \in \mathbb{R}^{p}, \mathbf{x}_{r} \in \mathbb{R}^{n-p}, \mathbf{A}_{1_p} \in \mathbb{R}^{m_{1} \times p},$ $\mathbf{A}_{1_r} \in \mathbb{R}^{m_{1} \times (n-p)}.$
Using this, Eq. (\ref{eq3p}) can be rewritten as:
\begin{equation}
\label{eq4g} 
\mathbf{y}=\left[\begin{matrix} \mathbf{y}_{p} \\ \mathbf{y}_{r} \end{matrix}\right]= \left[\begin{matrix} \mathbf{x}_{p} \\ \mathbf{x}_{r} \end{matrix}\right]+
\left[\begin{matrix} \mathbf{A}_{{E_{p}}}  \\  \mathbf{A}_{{E_{r}}} \end{matrix}\right]\mathbf{f}_{{E-1}}(...(\bm{\sigma}_{1}(\mathbf{A}_{{1_p}}\mathbf{x}_{p}+\mathbf{A}_{{1_r}}\mathbf{x}_{r}+\mathbf{b}_{{1}}))).
\end{equation}
The parts of the encoder of RAEO corresponding to the principal latent variables and residual latent variables are denoted by $\mathbf{e}_{p},\mathbf{e}_{r}$, respectively. Owing to the structure of the residual connection used in ResNet (see Fig. \ref{fig4a}) and identifying the residual latent variables with $\mathbf{V}_{\mathbf{y}_{r}}=\mathbf{0}$  results in: 
\begin{equation}
\label{eq4h} 
\begin{aligned}
\underbrace{\mathbf{x}_{r}+ \mathbf{A}_{{E_{r}}} \mathbf{f}_{{E-1}}(...(\bm{\sigma}_{1}(\mathbf{A}_{{1_p}}\mathbf{x}_{p}+\mathbf{A}_{{1_r}}\mathbf{x}_{r}+\mathbf{b}_{{1}})))}_{\mathbf{e}_{r}(\mathbf{x})} -\overline{\mathbf{y}}_{r}=\mathbf{0}
    \end{aligned}
\end{equation}
which represents an implicit relationship among variables in the data. Further, in the case of RAEO, it is possible to obtain an explicit relationship as well where the objective is to represent $\mathbf{x}_{r}$ as a nonlinear function of $\mathbf{x}_{p}.$ 
For that, select $\mathbf{A}_{{1_{r}}}=\mathbf{0}$ and retrain the RAEO to identify residual latent variables with $\mathbf{V}_{\mathbf{y}_{r}}=\mathbf{0}$ which modifies  Eq. (\ref{eq4h}) to an explicit relationship between $\mathbf{x}_{r}$ and $\mathbf{x}_{p}$ as
\begin{equation}
\label{eq4z}
   \mathbf{x}_{r}= \overline{\mathbf{y}}_{r} -\mathbf{A}_{{E_{r}}} \mathbf{f}_{{E-1}}(...\bm{\sigma}_{1}(\mathbf{A}_{{1_p}}\mathbf{x}_{p}+\mathbf{b}_{{1}})).
\end{equation}
This shows that the ResNet makes it easier to extract nonlinear relationships in explicit form. However, the success of getting an explicit relation cannot be guaranteed, i.e., achieving $\mathbf{V}_{\mathbf{y}_{r}}=\mathbf{0}$ with the constraint $\mathbf{A}_{{1_r}}=0$ may not always be possible. In such cases, one can explore rearranging the variables in $\mathbf{x}$ 
(so that the last $n-p$ variables can be represented as a function of the first $p$ variables) and retrain RAEO. 
\section{Theoretical results on AEO and RAEO}

This section presents theoretical results supporting the proposed AEO and RAEO. The main result is Theorem \ref{lemmaeo}, which shows that the global minimizer of the loss function in Eq. (\ref{eq3a}) guarantees variance ordering. The proof for Theorem \ref{lemmaeo} is based on the rearrangement inequality \cite{Hardy1934}, which is stated next.  

\begin{lemma}[Rearrangement inequality \cite{Hardy1934}]
   Let ${c_1, c_2, \dots, c_n}$ and ${d_1, d_2, \dots, d_n}$ be two sequences of real numbers arranged in non-decreasing order, i.e.,
\begin{equation}
c_1 \leq c_2 \leq \dots \leq c_n, \quad d_1 \leq d_2 \leq \dots \leq d_n.
\end{equation}
Then, for any permuted index set $\{p_1, p_2, \dots, p_n\}$   obtained by permuting the elements of the set $\{1,2,\dots,n\},$
the rearrangement inequality establishes that:
\begin{equation}
\label{eqrearr}
\begin{aligned}
\sum_{i=1}^{n} c_i d_{n - i + 1}
\leq \sum_{i=1}^{n} c_i d_{p_i}
\leq \sum_{i=1}^{n} c_i d_i.
\end{aligned}
\end{equation}
\end{lemma}
\begin{proof}
 Refer to \cite{Hardy1934}.    
\end{proof}
The next theorem presents a variance ordering guarantee of AEO and RAEO: 
\begin{theorem}[Variance ordering]
 \label{lemmaeo}
For any weighting matrix $\mathbf{Q}$ with entries satisfying $0 \leq q_1 < q_2 < \dots < q_m$, the global minimizers for AEO and RAEO obtained by solving Eq. (\ref{eq3y}) exhibit the following ordering in the latent variances:
\begin{equation}
\begin{aligned}
V_{y_1} \geq V_{y_2} \geq \dots \geq V_{y_m} \geq 0.
\end{aligned}
\end{equation}
\end{theorem}
\begin{proof}
    The proof proceeds by contradiction. Suppose that the global minimizer to Equation (\ref{eq3y}) results in an AEO in which the variances of the latent variables in $\mathbf{y}$ are not arranged in descending order. Let corresponding optimal loss function be (Eqs. \ref{eq3a},\ref{eq3y}):
\begin{align}
    \widehat{J} & = \widehat{J}_x + \alpha \widehat{J}_y + \beta \widehat{J}_{\theta}. 
\label{eqn:lossAEO}
\end{align}
Sort the latent variables of this AEO in descending order according to their variances, and then store the corresponding sorting indices in a vector denoted by $\mathbf{s}$. 
Next, we demonstrate that the loss associated with an AEO  exhibiting ordered latent variances is lower than the loss defined in Equation (\ref{eqn:lossAEO}). To establish this, we use the fact that permuting the latent variables as in Fig. \ref{figPAEO} does not affect the reconstruction error
where $\widetilde{\mathbf{y}}=\mathbf{P}\mathbf{y}$ is the permuted latent vector, $\mathbf{P}=\mathbf{I}_m(\mathbf{s},:)$  
is obtained by rearranging rows of the identity matrix $\mathbf{I}_m$ according to the sorting index vector $\mathbf{s}$, and  $\mathbf{y}=\mathbf{P}^{-1}\widetilde{\mathbf{y}}=\mathbf{P}^{-1}\mathbf{P}\mathbf{y}$.
With the permutation matrix $\mathbf{P}$, define an equivalent permuted-AEO (P-AEO):
\begin{equation} 
\begin{aligned}
&\widetilde{\mathbf{y}}=\widetilde{\mathbf{e}}(\mathbf{x})=\mathbf{P}\mathbf{e}(\mathbf{x})\\
&\widehat{\mathbf{x}}=\widetilde{\mathbf{d}}(\widetilde{\mathbf{y}})=\mathbf{d}(\mathbf{P}^{-1}\widetilde{\mathbf{y}}) 
\end{aligned}
\end{equation}
for which the loss function is
\begin{align}
    \widetilde{J} & = \widetilde{J}_x + \alpha \widetilde{J}_y + \beta \widetilde{J}_{\theta}. 
\label{eqn:lossP-AEO}
\end{align}
\begin{figure}[h]
 		\begin{center}
 		\includegraphics [scale=.475] {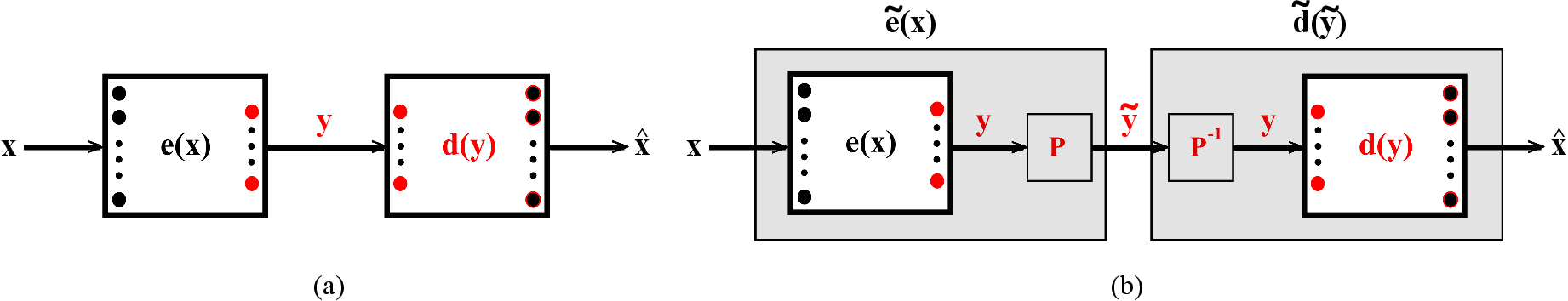}
 		\caption{{(a) AEO \hspace{0.3cm} (b) P-AEO.}}
 		\label{figPAEO}		
 	\end{center}
 \end{figure}
Since the reconstruction error and model parameters are identical for both the P-AEO and the original AEO, it follows that
\begin{align}
\widetilde{J}_{x} &= \widehat{J}_{x} \label{eqn:identicalSPE} \\
\widetilde{J}_{\theta} &= \widehat{J}_{\theta}.
\label{eqn:identicatparameterregularization}
\end{align} 
Further, the sample variance regularization term of P-AEO is:
\begin{align}   
\widetilde{J}_{y}&= Trace([\widetilde{\mathbf{Y}}-\overline{\widetilde{\mathbf{Y}}}]^{\top}\mathbf{Q}[\widetilde{\mathbf{Y}}-\overline{\widetilde{\mathbf{Y}}}]) \nonumber= ({N}-1)(q_{1}V_{y_{s_1}}+q_{2}V_{y_{s_2}}+ \dots+q_{m}V_{y_{s_m}}) \nonumber\\
& \leq ({N}-1)(q_{1}V_{y_1}+
\dots+q_{m}V_{y_m}) =  \widehat{J}_y
\label{eqn:rearrangementuse}
\end{align}
where inequality (\ref{eqn:rearrangementuse}) follows from rearrangement inequality (Lemma-1). Thus, from Eqs. (\ref{eqn:lossAEO}, \ref{eqn:lossP-AEO}-\ref{eqn:identicatparameterregularization}) and (\ref{eqn:rearrangementuse}), we obtain:
\begin{align}
\widetilde{J} & \leq \widehat{J}
\end{align}
since $\alpha,\beta>0$. This leads to a contradiction of the assumption that $\widehat{J}$ in Eq. (\ref{eqn:lossAEO}) is the optimal solution. The same proof applies to RAEO as well, since RAEO does not introduce any additional parameters compared to AEO.  Therefore, the optimal solution to Eq. (\ref{eq3y}) must correspond to an AEO (and RAEO) in which the latent variances are arranged in non-increasing order. This completes the proof.
\end{proof}

\section{Simulation results}
\subsection{Case study: Continuous stirred tank  reactor process}
\label{seccstraeo}
\begin{figure}[h]
 		\begin{center}
 		\includegraphics [scale=.475] {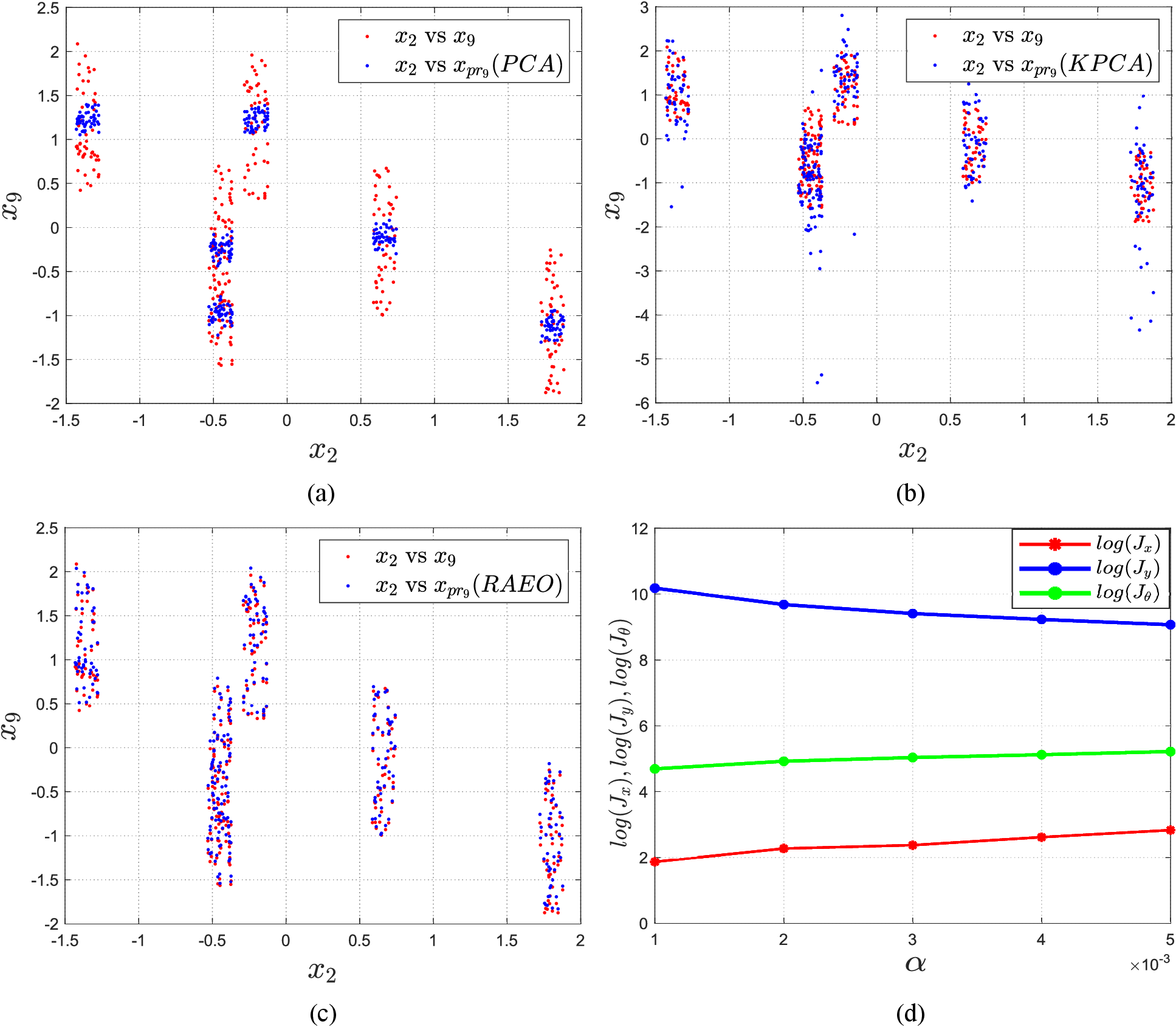}
 		\caption{{Model extraction (a) PCA, (b) KPCA,  (c) RAEO, (d) RAEO sensitivity analysis for $\alpha$.}}
 		\label{figcomp2}		
 	\end{center}
 \end{figure}

This section presents the application of the  AEO and RAEO in model extraction of a nonlinear continuous stirred tank reactor (CSTR) process. The CSTR steady-state process (excluding the linear constraints) consists of $n=10$ variables with 4 nonlinear equations as follows \cite{bGM24}: 
\begin{equation}
\label{eqsscstr}
    \begin{aligned}
        & \frac{x_{1}}{Ax_{10}} (x_{2}-x_{7}) - C_{d}x_{7} k_{0}  e^{-\frac{E}{Rx_{8}}} =0 \\
        & \frac{x_{1}}{Ax_{10}} (x_{3}-x_{8})+  \frac{1}{\rho C_{p}} C_{d}x_{7} k_{0}  e^{-\frac{E}{Rx_{8}}}(\varDelta H)  - \frac{U(x_{8}-x_{9})}{x_{10}\rho C_{p}} =0 \\
        &  \frac{x_{4}}{V_j} (x_{5}-x_{9}) + \frac{UA(x_{8}-x_{9})}{V_{j}\rho_{j} C_{p_j}} =0 \\
        & x_{7} k_{0}  e^{-\frac{E}{Rx_{8}}} A x_{10} - x_{6} =0.
    \end{aligned}
\end{equation}
The variables and constant parameters used in the simulation, along with their corresponding units and nominal values, are presented in Table \ref{table:Parameters}.
The training and testing data are generated by selecting $N=500$ samples  of $x_{1}$ to $x_{6}$, which are generated by adding random step changes to the nominal values of $x_{1},x_{2},x_{4}$ 
 (given in Table \ref{table:Parameters}), and remaining variables ($x_{3},x_{5},x_{6}$) are kept at their nominal values. Using this, the samples for $x_{7}$ to $x_{10}$ are generated by solving Eq. (\ref{eqsscstr}). Uniform noise with mean zero and standard deviation $1 \%$ of the nominal value is added to all the variables in the data to simulate measurement noise.  The first 300 samples are normalized and used as the training data, and the remaining are used as testing data. The parameters for AEO and RAEO are chosen as $E=2,D=2,$ $m_{1}=m_{3}=11,$ $m_{2}=m_{4}=n=10,$ $\alpha=0.01,$ $\beta=0.5$, and  the weighting matrix elements are selected exponentially: $q_{i}=c^{i}$ with $c=3.$ 
Table \ref{table:3} compares the performance of AEO and RAEO with PCA, KPCA, and AE which shows the variance of latent variables, the MSE for prediction error for variables $x_7$ to $x_{10}$ and the MSE for reconstruction error for variables $x_1$ to $x_{10}$. Here prediction error is defined as $x_i-x_{pr_i},~i=7,8,9,10$ with $x_{pr_i}$ being the solution of model equation as in Eq. \eqref{eqmodel} extracted using the various methods. 
The reconstruction error on the other hand is defined as $x_i-\hat{x}_i,~i=1,2,...,10$ where $\hat{x}_i$ is the value reconstructed by a method based on the reduced set of latent variables. 
In Table \ref{table:3}, the notations $\text{MSE}_{pr_{tr}}$ and $\text{MSE}_{pr_{ts}}$ denote the prediction errors, while $\text{MSE}_{re_{tr}}$ and $\text{MSE}_{re_{ts}}$ denote the reconstruction errors, on the training and testing datasets, respectively.
The simulation results show that the prediction errors with models extracted using AEO and RAEO are lower compared to PCA and AE. Fig. \ref{figcomp2}(a)-(c)  show the plot of $x_{2}$ $vs$ $x_{9}$ for both the actual and predicted data with PCA, KPCA, and RAEO, respectively. 
From
Fig. \ref{figcomp2}(a), it can be observed that the prediction by the PCA model cannot capture the full variation in the data.
Fig. \ref{figcomp2}(d)
plots the values of $J_{x},J_{y},J_{\theta}$ on a logarithmic scale for $\alpha$ varying from 0.001 to 0.005 in steps of 0.001, while $\beta=0.5$ and $c=3$ are held fixed. As illustrated in Figure \ref{figcomp2}(d), $J_{y}$ generally exhibits a decreasing trend as $\alpha$ increases, whereas $J_{x}$ and $J_{\theta}$ show increasing trends. 


\begin{table*}[t]
\centering
\caption{ Variables/constant parameters for CSTR process \cite{Bhushan2000}}
\begin{tabular}{c c c  c } 
 \hline
 Variable/Constants & Description & Value & Unit  \\ [0.5ex] 
 \hline\hline
  $x_{1}$ & Reactor inlet feed flow
 rate: $F_{i}$  &  40 & $\frac{ft^{3}}{h}$  \\ 
 $x_{2}$ & Inlet reactant concentration: $C_{Ai}$ & 0.50 &  lbmol of
 $\frac{A}{{ft^3}}$  \\
 $x_{3}$ & Inlet feed temperature: $T_{i}$  & 530 & \degree R \\
  $x_{4}$ & Coolant outlet flow rate: $F_{c}$  & 56.626 & $\frac{ft^3}{h}$  \\
   $x_{5}$ & Inlet coolant temperature: $T_{c_i}$  & 530 & \degree R \\
    $x_{6}$ & Vent flow rate: $F_{vg}$  & 10.6137 & $\frac{ft^3}{h}$ \\
     $x_{7}$ & Reactant concentration in reactor: $C_{A}$  & 0.2345 & lbmol of
 $\frac{A}{{ft^3}}$ \\
      $x_{8}$ & Reactor temperature: $T$  & 600 & \degree R \\
       $x_{9}$ & Jacket temperature: $T_{c}$  & 590 & \degree R \\
        $x_{10}$ & Height: $h$  & 0.32 & ft \\
    $C_d$ & Catalyst activity  & 1 & --  \\
    $k_{0}$ & Frequency factor  & $7.08\times 10^{10}$  & $h^{-1}$ \\
    $A$ & Heat-transfer area  & 150 & $ft^{2}$ \\
    $E$ & Activation energy  &  29900 & $\frac{btu}{lbmol}$ \\
    $R$ & Universal gas constant  & 1.99 &  \\
    $\rho$ & Density of process mixture  & 50 & $\frac{lbm}{ft^{3}}$ \\
    $\rho_{j}$ & Density of coolant  & 62.3 & $\frac{lbm}{ft^{3}}$ \\

    $C_{p}$ & Heat capacity (process side) & 0.75 & $\frac{btu}{lbm \degree R}$ \\
    $C_{p_j}$ & Heat capacity (coolant side)  & 1 & $\frac{btu}{lbm \degree R}$ \\
    $V_{j}$ & Volume of jacket  & 3.85 & $ft^{3}$ \\
    $U$ & Heat-transfer coefficient  & 150 & $\frac{btu}{h} ft^{2} \degree R$ \\
    $\varDelta H$ & Heat of reaction  & -30000 & $\frac{btu}{lbmol}$ \\
     $V$ & Volume of reactor  & 48 & $feet^{3}$ \\
 [1ex] 
 \hline
\end{tabular}
\label{table:Parameters}
\end{table*} 

\begin{table}[h!]
\centering
\caption{CSTR Example - Performance comparison.}
\begin{tabular}{c c c c c c} 
 \hline
Performance measure & PCA & KPCA & AE & AEO & RAEO   \\ [0.5ex] 
 \hline\hline
  $V_{y_1}$ & 4.4793  & - &   0.0102 &    0.3481 & 0.2698 \\ 
 $V_{y_2}$ & 1.4802  & - & 0.0553 & 0.2081 & 0.1105\\
 $V_{y_3}$ & 1.0756 & - & 0.3432 & 0.1126  &  0.0573\\
 $V_{y_4}$ &  1.0163 & - &0.0130 &  0.0497  & 0.0190\\
 $V_{y_5}$ & 0.9182 & - & 0.0016 & 0.0284  & 0.0131\\
 $V_{y_6}$ & 0.6771 & - &0.0072 & 0.0057  &  0.0059\\
 $V_{y_7}$ & 0.2219 & - &0.0061 &  0.0006  &   0.0001\\
 $V_{y_8}$ & 0.1244 & - & 0.0805& 0.0001  & 0.0001\\
 $V_{y_9}$ & 0.0062  & - &0.0222 & $1.1\times 10^{-5}$  & $8.1 \times 10^{-5}$\\
 $V_{y_{10}}$ & 0.0010 & - & 0.0028  & $1.6 \times 10^{-6}$  &  $4.5\times 10^{-5}$\\
 $\text{MSE}_{pr_{tr}}$  & 0.1230 & 6.5660 & 4.4897 & 0.0620 &  0.0027\\
$\text{MSE}_{pr_{ts}}$ & 0.2604 & 6.2390 & 0.8790 & 0.0138 & 0.1459\\
 $\text{MSE}_{re_{tr}}$ & 0.0352 &  0.0988 & 0.0001 & 0.1754 & 0.0383\\
 $\text{MSE}_{re_{ts}}$ & 0.0544 & 0.0944 & 42.7629 & 0.2486 & 0.1627\\
 [1ex] 
 \hline
\end{tabular}

\label{table:3}
\end{table}

\subsection{Application in real-time optimization (RTO)}

RTO deals with the problem of computing the optimal setpoint or steady-state operating point for industrial processes to improve efficiency,  product quality, process safety, etc \cite{Trierweiler2014}. The optimal setpoints are computed on a regular basis (daily, hourly, etc.) by solving a constrained,
steady-state optimization problem that requires the steady-state model of the process. The model extracted using AEO and RAEO can be used for solving RTO problems, which will be illustrated next on the CSTR system discussed in Section \ref{seccstraeo}. 
Here we use the model extracted with RAEO as the steady-state model, and our objective is to minimize the reactant concentration in the reactor (variable $x_{7}$ in Table \ref{table:Parameters}), since this results in maximum production. The resulting optimization problem for RTO is:
 \begin{equation}
    \begin{aligned}
&\underset{\mathbf{x}}{\min} \hspace{.2cm}  x_{7} \\
&\text{subj. to}  ~~ \mathbf{e}_{r}(\mathbf{x})-\overline{\mathbf{y}}_{r}=0\\
&\hspace{1.275cm} \mathbf{x}_{min} \leq \mathbf{x} \leq \mathbf{x}_{max}
     \end{aligned}
     \label{eqoptrto}
\end{equation} 
 where $\mathbf{e}_{r}(\mathbf{x})-\overline{\mathbf{y}}_{r}=0$ is the steady-state model identified using RAEO as in Eq. (\ref{eq4h}), $\mathbf{x}_{min},\mathbf{x}_{max} \in \mathbb{R}^{n}$ contain the minimum and maximum value of each variable in $\mathbf{x}$ which are defined as $90 \%$ and $110 \%$ of the nominal value, respectively. By solving the optimization problem in Eq. (\ref{eqoptrto}) the optimal value of $x_{7}$ is obtained as $0.2112$ for which the optimal setpoint becomes $\mathbf{x}^{*}=$ ${{\left[\begin{matrix}  39.9 & 0.4997 & 524.4
  & 56.5  & 534.5 & 10.7
  &  0.2112 & 605.6 & 599.9 & 0.3196\end{matrix}\right]^{\top}}}$.
    It can be observed that the optimal value of $x_{7}$ is not equal to $x_{7_{min}}=0.2110$ due to the constraints imposed by the model equations learned by RAEO.  This shows the application of the proposed methods in RTO. Further, the approaches can be used for other applications 
such as data reconciliation, process monitoring, etc.

\section{Conclusions}
This paper introduced a novel autoencoder framework, termed autoencoder with ordered variance (AEO), which incorporates a variance regularization term in the loss function to order the latent variables in decreasing variance. The architecture was further extended through the integration of residual connections, resulting in the ResNet AEO (RAEO). Both AEO and RAEO are designed to capture complex nonlinear relationships among variables in unlabeled datasets, enabling unsupervised model extraction. Theoretical analysis was presented to establish the variance ordering properties for both models. The proposed methods were illustrated through numerical examples, which highlight the superiority of AEO over PCA, KPCA, and AE in systems characterized by nonlinear relationships. 
The case study involving a CSTR process further illustrated the practical utility of the approach, particularly in a real-time optimization scenario. Future work will focus on extending the proposed framework to handle dynamic systems and time series data, thereby broadening its applicability to a wider range of real-world problems.

\appendix
\section{Appendix}
\subsection{ Linear model extraction using PCA}
\label{secA1}
PCA is an unsupervised dimensionality reduction method  \cite{Pearson1901,Jackson1991}.
Dimensionality reduction is achieved in PCA by linearly transforming the unlabeled data vector to the latent vector:
\begin{equation}
    \label{eq2s}\mathbf{y}=\mathbf{P}^{\top}\mathbf{x}
\end{equation}
where $\mathbf{x} \in \mathbb{R}^{n}$ is the data vector which is assumed to be mean centered, $\mathbf{y} \in \mathbb{R}^{m}$ is the latent vector, $ m \leq n,$  and matrix $ \mathbf{P} \in \mathbb{R}^{n\times m}$ contains $m$ orthonormal columns which are also called principal components. From $\mathbf{y},$ the reconstructed data vector is computed as 
\begin{equation}
\label{eq2t}
    \widehat{\mathbf{x}} = \mathbf{P}\mathbf{y}=\mathbf{P}\mathbf{P}^{\top}\mathbf{x}.
\end{equation}
where $\widehat{\mathbf{x}} \in \mathbb{R}^{n}$ is the reconstructed data vector.
One application of PCA is the identification of linear relationships among variables in a dataset. Such relationships can be inferred from latent variables exhibiting nearly zero variance, referred to as residual variables \cite{Narasimhan2008}. These variables satisfy the condition $V_{y_j} < \delta$, where $V_{y_j},$ $j=1,\dots,m$ denotes the variance of the latent variable $y_j$ and $\delta > 0$ is a tolerance value.
Since the latent variables in PCA are ordered in terms of decreasing variance, the residual variables will correspond to the last few elements of $\mathbf{y}$ with nearly zero variance. Consider $m=n$  and let the first $p$ latent variables have significant variance while the remaining $n-p$ are with negligible variance. Thus, 
 the first $p$ principal components are treated as significant and the rest $n-p$ as residual components, i.e., $\mathbf{P}=\left[\begin{matrix} \mathbf{P}_{p}  & \mathbf{P}_{r} \end{matrix}\right]$ where $\mathbf{P}_{p} \in \mathbb{R}^{n \times p}$ and $\mathbf{P}_{r} \in \mathbb{R}^{n \times (n-p)}.$
 Based on this, the latent variables can be partitioned into
\begin{equation}
    \label{eq2sa}
    \mathbf{y} =\left[\begin{matrix} \mathbf{y}_{p} \\ \mathbf{y}_{r} \end{matrix}\right] = \left[\begin{matrix} \mathbf{P}_{p} ^{\top} \\ \mathbf{P}_{r}^{\top} \end{matrix}\right]\mathbf{x}
\end{equation}
where $\mathbf{y}_{p}  \in \mathbb{R}^{p},$ $ \mathbf{y}_{r} \in \mathbb{R}^{(n-p)}$ contain the significant and residual variables.  The residual variables exhibit zero mean  and negligible variance, implying that $\mathbf{y}_{r} \approx 0$. Consequently, this approximation can be replaced by the equality $\mathbf{y}_{r} = 0$. Substituting this in Eq. (\ref{eq2sa}) results in:
\begin{equation}
    \label{eq2sb}
    \mathbf{P}_{r}^{\top}\mathbf{x}=\mathbf{0}
\end{equation}
which gives a linear model containing $n-p$ relationships amongst the $n$  variables.

\subsection{Nonlinear model extraction using Kernel PCA}
\label{secA2}
Kernel PCA (KPCA) is an extension of PCA for datasets that contain nonlinear relationships and is commonly used for nonlinear dimensionality reduction \cite{Scholkopf1998,Scholkopf1997}. The KPCA is based on mapping the data into high dimensional space where the relationships amongst features become linear, and hence PCA can be applied. Further the kernel trick is used to compute  inner product in feature space.  
Although the authors in \cite{Marimuthu2019} presented a kernel principal component regression–based approach for nonlinear model extraction and data reconciliation, the method requires the data to be labeled and therefore falls under a supervised learning framework.
 In this section, we discuss an unsupervised model extraction approach using KPCA, analogous to the PCA-based model extraction method in Section \ref{secA1}. In KPCA, the data vector $\mathbf{x}\in \mathbb{R}^{n}$ is mapped into high dimensional space using a nonlinear transformation which can be represented as:
\begin{equation}
\label{eqphi}
    \textbf{z}=\boldsymbol{\Phi}(\textbf{x})
\end{equation}
where  $\textbf{z}\in \mathbb{Z} \subseteq  \mathbb{R}^{h}, $ $ \boldsymbol{\Phi}:\mathbb{R}^{n} \rightarrow \mathbb{R}^{h}, $ and $ h>>n$.
Even if $\mathbb{Z}$ has arbitrarily large dimensionality, for certain
choices of $\boldsymbol{\Phi}$, we can still perform PCA in $\mathbb{Z}$ \cite{Scholkopf1998}. Let 
\begin{equation}
\mathbf{Z}=\left[\begin{matrix} \mathbf{z}_{1} & \mathbf{z}_{2} & \dots & \mathbf{z}_{N}\end{matrix}\right] \in \mathbb{R}^{h \times N}
\end{equation}
contain the feature samples which are assumed to be mean-centered. Thus covariance matrix in the feature space becomes:
\begin{equation}
\label{eqvz}
\textbf{V}_{\textbf{z}}=\frac{1}{N}\mathbf{Z}\textbf{Z}^{\top} \in \mathbb{R}^{h\times h}
\end{equation}
Applying eigenvalue decomposition on $\textbf{V}_{\textbf{z}}$ gives:
\begin{equation}
\label{eqvzu}
    \textbf{V}_{\textbf{z}} \textbf{u}=\lambda \textbf{u} \implies  \mathbf{Z}^{\top}\textbf{V}_{\textbf{z}} \textbf{u} = \lambda \mathbf{Z}^{\top} \mathbf{u}
\end{equation}
Further, the eigenvectors $\textbf{u} \in \mathbb{R}^{h}$ can be represented as linear combination of vectors in $\mathbf{Z}$ as \cite{Scholkopf1998}:
\begin{equation}
\label{eqzw}
    \textbf{u}=\textbf{Z}\textbf{w}
\end{equation}
where $\mathbf{w}\in \mathbb{R}^{N}$ are the coefficients. 
Substituting Eqs. (\ref{eqzw}) and (\ref{eqvz}) in Eq. (\ref{eqvzu}) results in:
\begin{equation}
    \begin{aligned}
        \label{eq_K2}\mathbf{Z}^{\top}\textbf{V}_{\textbf{z}} \textbf{Z}\textbf{w} = \lambda \mathbf{Z}^{\top} \textbf{Z}\textbf{w} \implies   \frac{1}{N}\underbrace{\mathbf{Z}^{\top}\textbf{Z}}_{\textbf{K}}\underbrace{\textbf{Z}^{\top} \textbf{Z}}_{\mathbf{K}}\textbf{w}=\lambda \underbrace{\mathbf{Z}^{\top} \textbf{Z}}_{\mathbf{K}}\textbf{w}
    \end{aligned}
\end{equation} 
where we defined the kernel matrix $\mathbf{K} \in \mathbb{R}^{N\times N}$ to be such that $\boldsymbol{\Phi}(\mathbf{x}_i)^{\top}
        \boldsymbol{\Phi}(\mathbf{x}_j)$ is its $(i,j)^{th}$ element. To obtain the solutions of Eq. (\ref{eq_K2}), we solve the eigenvalue problem:
\begin{equation}
\label{eq_kpca_evp}
\frac{1}{N} \mathbf{K}\textbf{w}=\lambda \textbf{w}
\end{equation}
for nonzero eigenvalues (See Appendix A of \cite{Scholkopf1998} for justification of this step). It is evident that every solution of Eq. (\ref{eq_kpca_evp}) is also a solution of Eq. (\ref{eq_K2}). 
           The elements of the kernel matrix $\mathbf{K}$ can be computed using the kernel functions without explicitly computing the inner product as:
\begin{equation}
    \boldsymbol{\Phi}(\mathbf{x}_i)^{\top}
        \boldsymbol{\Phi}(\mathbf{x}_j)=k(\mathbf{x}_i,\mathbf{x}_j)=\exp\left(
-\frac{\|\mathbf{x}_i - \mathbf{x}_j\|^2}{2\sigma^2}
\right)
\end{equation}
where $\sigma >0$ is the kernel width. Note that this corresponds to a specific kernel choice, namely the Gaussian kernel, whereas other kernel functions such as the polynomial kernel and the sigmoid kernel may also be employed \cite{Scholkopf1998}. In the kernel trick, instead of computing the eigenvectors of $\textbf{V}_{\textbf{z}}$ or $\textbf{Z}\textbf{Z}^{\top}$ we compute the eigenvectors of the kernel matrix $\textbf{K}$ or $\textbf{Z}^{\top}\textbf{Z}$ from which the eigenvectors of $\mathbf{V}_{\textbf{z}}$ can be computed using Eq. (\ref{eqzw}).  
To ensure zero-mean features, the kernel matrix is centered as
\begin{equation}
\mathbf{K}_{c}= 
\mathbf{K}
- \mathbf{1}_N \mathbf{K}
- \mathbf{K} \mathbf{1}_N
+ \mathbf{1}_N \mathbf{K} \mathbf{1}_N,
\end{equation}
where $\mathbf{1}_N = \frac{1}{N}\mathbf{1}\mathbf{1}^\top$ and $\mathbf{1}\in \mathbb{R}^{N}$ is a column vector containing all ones.
Kernel PCA is obtained by solving the eigenvalue problem
\begin{equation}
\mathbf{K}_c \textbf{w}_i = \lambda_i \textbf{w}_i,
\qquad i = 1,\dots,N,
\end{equation}
where $\lambda_i$ and $\textbf{w}_i$ denote the eigenvalues and
eigenvectors, respectively.
The eigenvectors are normalized according to
\begin{equation}
\textbf{w}_i \leftarrow
\frac{\textbf{w}_i}{\sqrt{\lambda_i}}.
\end{equation}
The first $p$ eigenvectors define the principal subspace, while the remaining
$N-p$ eigenvectors span the residual subspace:
\begin{equation}
\mathbf{W}_p =
\begin{bmatrix}
\textbf{w}_1 & \cdots & \textbf{w}_p
\end{bmatrix},\hspace{0.2cm}
\mathbf{W}_r =
\begin{bmatrix}
\textbf{w}_{p+1} & \cdots & \textbf{w}_N
\end{bmatrix}.
\end{equation}
Then similar to PCA, model extraction can be achieved in the residual space as:  
\begin{equation}
\mathbf{W}_r^\top \mathbf{k}_{c}({\mathbf{x}}) = \mathbf{0},
\end{equation}
where
\begin{equation}
\mathbf{k}({\mathbf{x}})
=
\begin{bmatrix}
k({\mathbf{x}},\mathbf{x}_1) &
\cdots &
k({\mathbf{x}},\mathbf{x}_N)
\end{bmatrix}^\top
\end{equation}
and $\mathbf{k}_c({\mathbf{x}})$ denotes the centered kernel vector.
The resulting system of nonlinear equations can be solved using a numerical
root-finding method.

\section*{Acknowledgments}
This work was supported by the Science and Engineering Research Board, Department of Science and Technology India, through grant number CRG/2022/002587.

\bibliographystyle{plain}

\begin{thebibliography}{1}

\bibitem{Cayton2005}
L. Cayton, \emph{\textquotedblleft Algorithms for manifold learning,\textquotedblright} Technical report CS2008-0923, University of California, 2005.


\bibitem{Pearson1901}
K. Pearson, \emph{\textquotedblleft On lines and planes of closest fit to systems of points in space,\textquotedblright} 
 Philosophical magazine, vol. 2, pp. 559-572, 1901.

 \bibitem{Jackson1991}
J. Jackson, \emph{\textquotedblleft A  user’s guide to  principal components,\textquotedblright} 
Wiley, New York, 1991.


\bibitem{Narasimhan2008}
S. Narasimhan and
 S. Shah, \emph{\textquotedblleft Model identification and error covariance matrix estimation
from noisy data using PCA,\textquotedblright} 
Control Engineering Practice, vol. 16, pp. 146–155, 2008.


\bibitem{Kramer1989}
M. Kramer, \emph{\textquotedblleft Nonlinear principal component analysis using 
autoassociative neural networks,\textquotedblright} 
AIChE Journal, vol. 37, pp. 233-243, 1989.




\bibitem{Hastie1989}
T. Hastie and W.  Stuetzle,\emph{\textquotedblleft Principal curves,\textquotedblright} 
Journal of the American Statistical Association, vol. 84, pp. 502-516, 1989.





\bibitem{Scholkopf1998}
B. Scholkopf, A. Smola, and K. Muller,  \emph{\textquotedblleft Nonlinear component analysis
as a kernel eigenvalue problem,\textquotedblright} 
Neural Computation, vol. 10, pp. 1299–1319, 1998.

\bibitem{Tenenbaum2000}
J. Tenenbaum, L. Silva, \emph{\textquotedblleft A global geometric framework for nonlinear dimensionality reduction,\textquotedblright} Science, vol. 290, pp. 2319–2323, 2000.




\bibitem{Nadler2005}
B. Nadler, S. Lafon, R. Coifman, I. Kevrekidis, \emph{\textquotedblleft  Diffusion maps, spectral clustering, and eigenfunctions of
fokker–planck operators,\textquotedblright} $19^{th}$ Conference on
Neural Information Processing Systems, Vancouver, Canada, 2005.


\bibitem{Li2022}
P. Li, Y. Pei, and J. Li, \emph{\textquotedblleft A comprehensive survey on design and application of autoencoder in deep learning,\textquotedblright} 
Applied Soft Computing, vol. 138, pp. 110-176, 2022.


\bibitem{Kingma2014}
D. Kingma,  and M. Welling, \emph{\textquotedblleft Auto-encoding variational
bayes,\textquotedblright} 
International Conference on Learning Representations, 2014.


\bibitem{Digeil2022}
M. Digeil, Y. Grinberg, D. Melati, M. Dezfouli, J. Schmid, P. Cheben, S. Janz, and D. Xu, \emph{\textquotedblleft PCA-boosted autoencoders for nonlinear dimensionality reduction in low data regimes,\textquotedblright} 
arXiv, 2022.

\bibitem{bSR}
S. Rifai, P. Vincent, X. Muller, X. Glorot, and Y. Bengio, \emph{\textquotedblleft Contractive autoencoders:
explicit invariance during feature extraction,\textquotedblright} 
28$^{th}$ International Conference on Machine Learning, Bellevue, WA, USA, 2011.

\bibitem{Zhao2018}
H. Zhao,\emph{\textquotedblleft Neural component analysis for fault detection,\textquotedblright} 
Chemometrics and Intelligent Laboratory Systems, vol. 76, pp. 11-21, May 2018.

\bibitem{Wang2019}
W. Wang, D. Yang, F. Chen, Y. Pang, S. Huang, and Y. Ge, \emph{\textquotedblleft Clustering with orthogonal autoencoder,\textquotedblright} 
IEEE Access, vol. 7, pp. 62421-62432, 2019.

\bibitem{Pham2022}
C. Pham, S. Ladjal, and A. Newson, \emph{\textquotedblleft PCA-AE: principal component analysis - autoencoder for organizing the
latent space of generative networks,\textquotedblright} 
Journal of Mathematical Imaging and Vision, vol. 64, pp. 569–585, Apr. 2022.


\bibitem{Trunz2022}
E. Trunz, M. Weinmann, S. Merzbach, R. Klein, \emph{\textquotedblleft Efficient structuring of the latent space for controllable data reconstruction and compression,\textquotedblright} 
Graphics and Visual Computing,
vol. 7, pp. 1-13, Dec. 2022.





\bibitem{Hinton2006}
G. Hinton and R. Salakhutdinov, \emph{\textquotedblleft Reducing the dimensionality of data with neural networks,\textquotedblright} 
Science, vol. 73, pp. 504-507, 2006.


\bibitem{He2016}
K. He, X. Zhang, S. Ren, and J. Sunu, \emph{\textquotedblleft Deep residual learning for image recognition,\textquotedblright} 
IEEE Conference on Computer Vision and Pattern Recognition, pp. 770-778, 2016.

\bibitem{Wickramasinghe2021}
C. Wickramasinghe, D. Marino, and M. Manic, \emph{\textquotedblleft ResNet autoencoders for unsupervised feature
learning From high-dimensional data: deep
models resistant to performance degradation,\textquotedblright} 
IEEE Access, vol. 9, pp. 40511-40520, 2021.













\bibitem{Benton2018}
D. Benton,\emph{\textquotedblleft Nonlinear equations: Numerical methods for solving,\textquotedblright} 
Amazon Digital Services LLC - Kdp Print USA, 2018.

\bibitem{Marimuthu2019}
K. Marimuthu and S. Narasimhan, \emph{\textquotedblleft Nonlinear model identification and data reconciliation using kernel
principal component regression,\textquotedblright} 
Industrial and Engineering Chemistry Research, vol. 58, pp. 11224-11233, 2019.

\bibitem{Shang2014}
C. Shang, F. Yang, D. Huang, and W. Lyu, \emph{\textquotedblleft Data-driven soft sensor development based on deep learning
technique,\textquotedblright} 
Journal of Process Control, vol.  24, pp. 223–233, Feb. 2014.


\bibitem{Lee2004}
J. Lee, C. Yoo, S. Choi, P. Vanrolleghem, and I. Lee, \emph{\textquotedblleft Nonlinear process monitoring using kernel principal component analysis,\textquotedblright} 
Chemical Engineering Science, vol.  59, pp. 223 – 234, Jan. 2004.


\bibitem{ZLi2022}
Z. Li, L. Tian, Q. Jiang, and X. Yan, \emph{\textquotedblleft Dynamic nonlinear process monitoring based on
dynamic correlation variable selection and kernel
principal component regression,\textquotedblright} 
Journal of the Franklin Institute, vol. 359, pp. 4513-4539, Jun. 2022.



\bibitem{Trierweiler2014}
J. Trierweiler, \emph{\textquotedblleft Real-time optimization of industrial processes,\textquotedblright} 
Encyclopedia of Systems and Control, pp. 1 – 11, Jan. 2014.
{\bibitem{Hardy1934} G. Hardy, J. Littlewood, and G. Polya,  \emph{\textquotedblleft Inequalities,\textquotedblright} 
Cambridge University Press, 1934.}


\bibitem{bGM24}
G. Patel and M. Bhushan, \emph{\textquotedblleft 
Reliability-based sensor placement design in nonlinear processes using cumulative residual Kullback-Leibler divergence,\textquotedblright} 
Industrial $\&$ Engineering Chemistry Research, vol. 63, pp. 9869 – 9886, May 2024.

\bibitem{Bhushan2000}
M. Bhushan and R.  Rengaswamy \emph{\textquotedblleft 
Design of sensor network based on the signed directed graph of the process for efficient fault diagnosis,\textquotedblright} 
Industrial $\&$ Engineering Chemistry Research, vol. 19, pp. 999-1019, Mar. 2000.


\bibitem{Scholkopf1997}
B. Scholkopf, A. Smola, and K. Muller,  \emph{\textquotedblleft Kernel principal component analysis,\textquotedblright} 
Artificial Neural Networks ICANN, 1997.



\end{thebibliography}

\vfill

\end{document}